\documentclass[12pt]{revtex4}
\usepackage{amsfonts}

\begin{document}

\large
\noindent 
{\bf Domain-averaged Fermi hole versus regional reduced density matrices:
    a critical comparison} 

\vskip 10mm 
\normalsize

\noindent
{\bf Diego R. Alcoba$^{a}$, Roberto C. Bochicchio$^{a*}$, 
Luis Lain$^{b}$, Alicia Torre$^{b}$}

\vskip 40mm

\noindent 
In their recent work Cooper and Ponec [Phys. Chem. Chem. Phys., 2008, 10, 1319-1329]
proposed a ``one-electron approximation to domain-averaged Fermi hole (DAFH)" used in 
electronic population studies. The goal of this comment is to note that the proposal had 
been already published within the framework of domain-restricted reduced density matrices 
($\mathrm{\Omega}$-RDM) and to show that it cannot conceptually be considered
as an approximation to DAFH as the authors invoke.

\vskip 50mm
\small
\noindent
\rule{155mm}{0.2mm}
\noindent
{\small
$^{a}${\it Departamento de F\'\i sica, Facultad de Ciencias Exactas y
Naturales, Universidad de Buenos Aires, Ciudad Universitaria, 1428,
Buenos Aires, Argentina.} {\it E-mail: rboc@df.uba.ar}\\
$^{b}${\it Departamento de Qu\'{\i}mica F\'{\i}sica. Facultad de Ciencias.
Universidad del Pa\'{\i}s Vasco. Apdo. 644 E-48080 Bilbao, Spain.}

\newpage

\normalsize 
\noindent 
The {\it domain-averaged Fermi hole} (DAFH) approach to analyze electronic 
structures has been widely used in the recent past$^{1}$. However, its 
correlated counterpart (correlated hole) has not been considered on an equal 
footing due to the high computational cost required for the use of  
second-order reduced density matrices (2-RDM), $^{2}$D. In ref. 1, the 
authors mention their early work using these matrices$^{2}$ but omit to 
advise that the first attempt to extend domain-averaged Fermi hole analysis 
to the correlated case has been published by some of us in ref. 3, making 
explicit use of the $^{2}$D structure in matrix form. The aim of the authors
of ref. 1 is to approach the DAFH analysis by a simple model that avoids the 
use of $^{2}$D. The goal of this report is twofold. On the one side 
to show that this one-electron model the authors claim for themselves in 
this reference is nothing but a model previously developed by us within the 
scenario of the domain-restricted first-order reduced density matrices$^{4}$, 
$^{1}\mathrm{D(\Omega)}$. On the other side we attempt to clarify the physical 
and mathematical reasons, misunderstood in ref. 1, showing that what is there
called one-electron approximation to DAFH is unsuitable for such a task.

Let us first show that the approach called {\it pseudo}-DAFH 
[cf. eqn (2.9) of ref. 1] is nothing but the model we called {\it
symmetric} approach to $^{1}\mathrm{D(\Omega)}$ in ref. 4. Following the 
authors in ref. 1, we will focus attention on closed-shell wave functions. 
Eqn (11) in ref. 4 reads, 

\begin{equation}
^{1}\mathrm{D}^{i}_{j}(\Omega) =\; \sum_{k,l}\;
(^{1}\mathrm{D}^{\frac{1}{2}})^{i}_{k}\;\mathrm{S}(\Omega)^{k}_{l}
(^{1}\mathrm{D}^{\frac{1}{2}})^{l}_{j}
\label{1}
\end{equation}
\vskip 2mm 

\noindent 
where $(^{1}\mathrm{D}^{\frac{1}{2}})^{i}_{j}$ mean the elements of the positive 
square root matrix arising from the spin-free first-order reduced density matrix 
(1-RDM), $^{1}\mathrm{D}$, and $\mathrm{S}(\Omega)^{i}_{j}= <i|j>_{\Omega}$ 
(${a^{i}_{j}}^{\Omega}$ in ref. 1) are the overlap integrals over Bader domains 
$\Omega$ in an orthonormal basis set labeled as $i,j, k,l, ....$ $^{4}$. This 
$^{1}\mathrm{D}(\Omega)$ arises from a partitioning of $^{1}\mathrm{D}$ in terms 
of Bader's regions, so that $^{1}\mathrm{D}^{i}_{j} = \sum_{\Omega}\;
^{1}\mathrm{D}^{i}_{j}(\Omega)$. The expression of eqn (1) in the natural basis 
set in which ${^{1}\mathrm{D}}^{i}_{j}=n_{i} \delta_{i}^{j}$ 
where $n_{i}$ and $\delta_{i}^{j}$ stand for the natural occupation numbers and 
the Kronecker delta, respectively, yields the symmetric form 
$\mathrm{G}^{\Omega}\; =\; {n}^{\frac{1}{2}}\; \mathrm{S}(\Omega)\; 
{n}^{\frac{1}{2}}$ of ref. 1 [cf. eqn (2.9)]. Let us note that eqn (1) formulates 
a true domain-restricted first-order reduced density matrix associated 
with the domain $\Omega$, i.e., a representable matrix, $^{1}\mathrm{D}(\Omega)$$^{5}$, 
which is valid for any type of wave function$^{5}$, such as independent particle 
models and correlated particle models with non-integer occupation numbers$^{5}$. 

Our second aim is to show the inconsistencies of the proposal of one-electron
approximation to DAFH in ref. 1. The errors arise from wrong assumptions
of mathematical and physical nature performed to state the there called 
{\it pseudo}-DAFH$^{1}$ model as an approximation to DAFH. Let us first explain 
the scenario of the domain-restricted decompositions of the 1-RDM, in order to 
analyze these assumptions. A reduced density matrix must fulfil necessary 
and sufficient conditions to be representable, i.e., to assure that there 
exists a wave function or a statistical ensemble, though unknown, from which 
it derives. The $1$-RDM is hermitian, positive semidefinite and bounded$^{5}$ 
and is normalized to the number of electrons, N. The necessary and sufficient 
conditions to be fulfilled by a closed-shell system domain-restricted 
first-order density matrix $^{1}\mathrm{D}(\Omega)$ (for any domain $\Omega$ 
in the system) are that its eigenvalues, $n_{i}^{\Omega}$ must lie within the 
real interval $[0,2]$, i.e., $0 \le n_{i}^{\Omega} \le 2\;$$^{5}$; 
note that the domains $\Omega$ define open systems which necessarily become 
described by grand-canonical ensembles$^{5}$. Let us now consider the one-electron 
approximation to DAFH in ref. 1. This approximation is nothing but our symmetric 
approach to $^{1}\mathrm{D}(\Omega)$ as we have shown above. Consequently, 
it turns out to be a true domain-restricted first-order density matrix, 
according to the enumerated representability conditions. An appropriate model 
must predict approximate values of a determined quantity keeping its physical 
nature; that is not the case of the symmetric model of $^{1}\mathrm{D}(\Omega)$ 
in relation with the approximation of DAFH matrices. Thus, as has been shown, 
both local$^{7}$ and non-local (or integrated) matrix formulations$^{3}$ of 
the correlated hole (DAFH) are not positive semi-definite$^{2,3}$. Consequently, 
negative eigenvalues (populations) can appear from both formulations, which make 
this quantity physically unacceptable as a density. Furthermore, populations 
greater than $2$ do not fulfilling the Pauli principle arise. Examples showing 
populations out of the interval $[0,2]$ may be found in both the local and 
domain-averaged formulations of correlated hole$^{3,7}$. Simple systems like 
$\mathrm{N}_{2}$ molecule at equilibrium geometry calculated at a single-double 
configuration interaction level with the PSI 3.2 package$^{8}$ in a 6-31G basis 
set, exhibit negative eigenvalues ($\sim -0.03$)$^{9}$. This is a significant 
negative value which must be neglected in DAFH analysis, as was made in refs. 
1-3 and consequently the isopycnic transformation$^{10}$ used to localize the 
eigenvectors of DAHF in those references is no longer valid. Also, to neglect 
the negative eigenvalues of the DAFH matrices permits the density to be 
delocalized due to the particle conservation, i.e., the density integration 
over a domain does not keep the right population for the domain and thus for 
the whole system$^{1}$. This is not the case of the symmetric model of 
$^{1}\mathrm{D}(\Omega)$ which do not behave in this way, since it can properly 
support an isopycnic transformation because all its eigenvalues are positive. 
Thus, this difficulty is completely avoided within the framework of 
domain-restricted first-order reduced density matrix theory, which provides 
a more localized picture of electron distributions as shown in the comparison 
of both $^{1}\mathrm{D}(\Omega)$ and DAHF pictures$^{4}$. These results may not 
be considered as unexpected because, as shown above, correlated DAFH are not 
true {\it particle} densities. Namely, DAFH may not satisfy the rigorous 
conditions to be a {\it true} first-order reduced density matrix, and thus 
it may not properly describe an open system. Consequently, DAFH are not density 
matrices but different entities and it is a severe conceptual shortcoming to 
approximate DAFH by means of any approach to $^{1}\mathrm{D}(\Omega)$ including 
the symmetric one. In fact, the above commented results induced us to study 
decomposition schemes in the physical space providing the 
$^{1}\mathrm{D}(\Omega)$ matrices$^{4,6}$. 

It may be noted that DAHF model turns out to be a true density matrix only 
for closed-shell wave functions having $2$ or $0$ orbital populations. In this
case it is equivalent to our symmetric approach because then the density 
cumulant terms of $^{2}\mathrm{D}$ vanish and $^{1}\mathrm{D}$ is 
{\it duodempotent} (${^{1}\mathrm{D}}^{2}=2\; {^{1}\mathrm{D}}$)$^{4}$. Other 
important remark to be made in relation with the topics treated in ref. 1 is 
that this one-electron symmetric approach $^{1}\mathrm{D}(\Omega)$ to DAFH has 
not a common basis of eigenvectors for all the $\Omega$ domains in the system, 
i.e., the corresponding matrices are not diagonal in the same basis set$^{6}$. 
Therefore, calculating quantum chemical descriptors depending on the eigenvalues 
and eigenvectors of two domains, as made in refs. 1, 2 and 4, is neither 
mathematically nor physically rigorous$^{6}$. This drawback may be avoided 
using a model with a common eigenbasis to all domains such as the isopycnic 
approach to the domain-restricted decomposition of the first-order reduced 
density matrix$^{6}$.

\vskip 5mm

\noindent
Grants in aid from the Universidad de Buenos Aires, (Project No. X-024), the Consejo 
Nacional de Investigaciones Cient\'{\i}ficas y T\'enicas, Rep\'ublica Argentina 
(PIP No. 5098/05), the Spanish Ministry of Education (Grant No. CTQ2006-01849/BQU) 
and the Universidad del Pais Vasco (Grant No. GIU06/03) are acknowledged. 

\vskip 10mm

\noindent
{\bf References}
\vskip 3mm
\small
\noindent
\hspace*{1mm} 1 \hspace*{1mm} D. L. Cooper and R. Ponec, 
{\it Phys. Chem. Chem. Phys.}, 2008, {\bf 10}, 1319-1329 and\\ \hspace*{7mm} 
references therein.

\noindent
\hspace*{1mm} 2 \hspace*{1mm} R. Ponec and D. L. Cooper, {\it Faraday Disccuss.}, 2007, 
{\bf 135}, 31-42.

\noindent
\hspace*{1mm} 3 \hspace*{1mm} R. C. Bochicchio, L. Lain and A. Torre, {\it J. Chem. Phys.}, 
2005, {\bf 122}, 084117.

\noindent
\hspace*{1mm} 4 \hspace*{1mm} D. R. Alcoba, L. Lain, A. Torre and R. C. Bochicchio, 
{\it J. Chem. Phys.}, 2005, {\bf 123},\\ \hspace*{7mm} 144113.

\noindent
\hspace*{1mm} 5 \hspace*{1mm} D. R. Alcoba, R. C. Bochicchio, G. E. Massacessi, L. Lain 
and A. Torre, {\it Phys. Rev. A},\\ \hspace*{7mm} 2007, {\bf 75}, 012509.

\noindent
\hspace*{1mm} 6 \hspace*{1mm} D. R. Alcoba, R. C. Bochicchio, A. Torre and L. Lain, 
{\it J. Phys. Chem. A}, 2006, {\bf 110},\\ \hspace*{7mm} 9254-9260.

\noindent
\hspace*{1mm} 7 \hspace*{1mm} M. A. Buijse, {\it Electron Correlation}, Centrale 
Huisdrukkerij Vrije Universiteit, Amsterdam,\\ \hspace*{7mm} 1991.

\noindent
\hspace*{1mm} 8 \hspace*{1mm} T. D. Crawford, C. D. Sherrill, E. F.  Valeev, 
J. T.  Fermann, R. A. King, M. L. Leininger,\\ \hspace*{7mm} S. T.  Brown, 
C. L. Janssen, E. T. Seidl, J. P. Kenny and W. D. Allen, PSI 3.2 2003.

\noindent
\hspace*{1mm} 9 \hspace*{1mm} D. R. Alcoba, R. C. Bochicchio, L. Lain and 
A. Torre, unpublished results. 

\noindent
10 \hspace*{1mm} J. Cioslowski, {\it Int. J. Quantum Chem.}, 1990, {\bf S24}, 15.

\end{document}